\documentclass[11pt,twoside]{article}
\usepackage{asp2004}
\usepackage{psfig}
\usepackage{epsf}
\usepackage{graphics}
\usepackage{lscape}
\def\edcomment#1{\iffalse\marginpar{\raggedright\sl#1\/}\else\relax\fi}
\reversemarginpar

\begin{document}
\title{Direct Evidence for Dynamical Evolution of Luminosity Functions
  of Globular Cluster Systems: HST/ACS observations of the 3-Gyr-old merger
  remnant NGC 1316}
\author{Paul Goudfrooij}
\affil{Space Telescope Science Institute, 
  Baltimore, Maryland, 
  USA}

\begin{abstract}
Recent observations of globular clusters
(GCs) in intermediate-age (2--4 Gyr old), early-type merger remnants 
have provided the hitherto `missing link' between young merger remnants and
`normal' elliptical galaxies in the form of a GC subsystem with colors and
luminosities consistent with population synthesis model predictions
for those ages and $\sim$\,solar metallicity. 
Here we present new, 
deep observations of the GC system of the intermediate-age merger 
remnant NGC~1316, using the {\it ACS} camera aboard {\it Hubble Space
  Telescope}, which allowed us to create luminosity functions
(LFs) {\it as a function 
  of galactocentric radius}. We find that the inner 50\% of the `red' GC
system shows a clear turnover in its LF, at about 1 mag fainter than that of
the `old' blue GCs. This constitutes direct, dynamical evidence that
metal-rich GC populations formed during a gas-rich merger can evolve into the
`red', metal-rich GC populations that are ubiquitous in `normal' giant
ellipticals.  
\end{abstract}
\thispagestyle{plain}

\section{Galaxy Mergers and Globular Cluster Formation}

Over the last decade, the importance of young massive star clusters as
signposts to major star formation events has been recognized. In particular,
the discovery of young globular clusters (GCs) in galactic mergers and young
merger remnants using {\it Hubble Space Telescope \/} (HST) imaging (e.g.,
Schweizer 2002 and references therein) has generated a great
deal of excitement. This is hardly surprising, as these clusters have provided
previously unexpected opportunities including {\it (i)\/} the chance to study
the formation and evolution of GCs in the local universe rather than trying
to derive how they formed $\sim$\,14 Gyr ago; {\it (ii)\/} the possibility to
age-date merger events (i.e., major star formation epochs) in galaxies from
star clusters with simple, single-burst stellar populations. 

In many cases, follow-up spectroscopy of these young clusters confirmed their
nature as star cluster and their  
ages, and in one case even their high masses predicted from their colors and
luminosities (Maraston et al.\ 2004). The metallicities of the young clusters
have been found to be near solar, as expected for clusters formed out of 
enriched gas in spiral disks. Such metal-rich clusters 
are now known in merger remnants at essentially all ages, ranging from birth
(e.g., NGC~4038/4039, Whitmore et al.\ 1999) through youth (several 10$^8$ yrs,
e.g.\ NGC 7252, Schweizer \& Seitzer 1998) to middle age, when only faint
ripples and loops reveal their merger history (e.g. NGC~1316,
Goudfrooij et al.\ 2001a). 

It should also be emphasized that the cluster systems of young merger
remnants were also found to contain fainter, redder objects with colors and
luminosities consistent with those of old, metal-poor halo GCs that are very
common in nearby galaxies such as the Milky Way. It is commonly
believed that these fainter GCs in merger remnants probably belonged to the
(now merged) progenitor spirals.   
This bimodality in star cluster systems of young merger remnants is rather
interesting in the context of formation scenarios for {\it normal\/}
elliptical galaxies, since the latter are also well known to generally host
GC systems with bimodal color distributions 
(Kundu \& Whitmore 2001; Larsen et al.\ 2001 to name a few).  
Deep {\it HST\/} imaging of `normal' ellipticals shows that roughly
half of their GCs are blue (metal-poor) and half are red (metal-rich,
typically roughly solar or somewhat less). Spectroscopy with 10-m class
telescopes revealed that both `blue' and `red' GC subpopulations of
`normal' ellipticals are typically old ($\ga$\,8 Gyr, 
Cohen, Blakeslee \& C\^ot\'e 2003; Puzia et al.\ 2004). A natural
interpretation of these data is that the metal-rich GCs in 
`normal' ellipticals formed in gas-rich mergers at $z \ga 1$, and that the
formation process of galaxies with significant populations of metal-rich GCs
was similar to that in galaxy mergers observed today. 

One might think ``now don't get carried away'' at this point, and indeed it is
important to examine this scenario in 
detail. If correct, one {\it should\/} be able 
to {\it (i)\/} find ellipticals with second-generation GCs of intermediate
age (i.e., 2\,--\,5 Gyr) and {\it (ii)\/} study the evolution of
second-generation GC systems from young through intermediate to old ages,
and compare their properties with theoretical predictions. These two issues
are addressed in the remainder of this paper. 

\section{Intermediate-Age Globular Cluster Systems}

The existence of intermediate-age GC systems has been argued for several
years already, based on GC colors and luminosities from deep optical
photometry of ellipticals with high fine structure such as ripples and tidal
features (e.g., Whitmore et al.\ 1997). 
The breakthrough in identifying intermediate-age GCs came by using methods
that break the age-metallicity degeneracy present in optical colors, namely 
the use of optical-to-near-IR colors and spectroscopy with large
telescopes. Goudfrooij et al.\ (2001a,b) used both methods to discover a
major $\sim$\,3 Gyr old, metal-rich GC population in the merger remnant
NGC~1316, discussed further below. Soon afterward, Puzia et al.\ (2002)
identified a significant population of intermediate-age GCs in the elliptical
galaxy NGC~4365 (featuring a counter-rotating core) using $V$, $I$, and
$K$-band photometry. Follow-up Keck spectroscopy of a subset of the Puzia et
al.\ GCs by Larsen et al.\ (2003) confirmed the effectiveness of the
optical-to-near-IR technique, which was subsequently used by Hempel et al.\
(2003) who identified intermediate-age GCs in NGC~5846 as well. Other
investigations using this method are ongoing.  

Now that the presence of intermediate-age GC systems has been established,
time has come to test whether or not properties of GC systems formed in
mergers are compatible with those of `red' GCs in old ellipticals. If so, this
would render them a definite evolutionary link between young remnants and
`normal' ellipticals with bimodal GC color distributions. One of the most
important tracers of such systematic evolution is the luminosity function
(LF) of second-generation GCs. The LF of `old' GC systems is a Gaussian in
magnitude units, with a peak (`turnover') at $M_V = -7.3$ 
(e.g., Harris 1996), while that of GCLF's in young merger remnants is a
power law with index $\alpha \sim -2$ (e.g., Whitmore et al.\ 1999). The
transition from power law to Gaussian LF has been predicted theoretically (Fall
\& Zhang 2001), being a consequence of preferential erosion of low-mass
GCs due to various disruption mechanisms, of which the main ones are
internal two-body relaxation and tidal shocking. {\it If indeed the
  metal-rich GCs in `old' giant ellipticals are evolved GCs formed during a
  gas-rich merger, this gradual erosion should be evident in observed LFs of
  second-generation GC systems that form an age sequence}. This critical
test of the `merger hypothesis' was infeasible until recently, due to the
necessity to reach $\sim$\,2 mag beyond the turnover magnitude. 
However, the unprecedented sensitivity of the {\it ACS\/} camera, installed
on {\it HST\/} in March 2002, made this possible.  

\section{Results}

We observed NGC~1316 in March 2003 using the wide-field channel of {\it
  HST/ACS\/} through the F435W, F555W, and F814W filters, with total exposure
times of 1860 s, 14560 s, and 4770 s, respectively. 
The supreme spatial resolution of
the new {\it ACS\/} images allowed us to impose stringent {\it size
  constraints\/} on the list of detected compact objects so as to exclude
extended background galaxies as well as foreground stars. Due to space
restrictions, I will only show the main result here: {\it An observational
  confirmation of the predicted dynamical evolution of the LF of a
  second-generation, metal-rich GC system}. 
A paper containing a more detailed presentation and discussion of the results 
of this study has been submitted to a refereed journal.  

The new {\it ACS\/} images allowed us to reach a factor of $\sim$\,4.5
fainter in luminosity than with the WFPC2 observations presented in
Goudfrooij et al.\ (2001b; hereafter Paper I), which resulted in a final GC
candidate list of 1496 objects - almost four times as many as from the WFPC2
images (!).  
Metal-poor (blue) GCs were selected using $V\!-\!I < 0.95$ (just
redward of the reddest halo GC in our Galaxy), whereas metal-rich (red) GCs
were selected using $V\!-\!I > 1.05$. Correction for contamination by
{\it compact\/} background galaxies was done in a statistical way, using 
WFPC2 F555W and F814W images that were taken in parallel with the {\it ACS\/}
data. After this correction, the LFs of the blue and red subpopulations were
produced: See panels (a) and (b) of Fig.\ 1. The blue GC system shows a
Gaussian LF consistent with that of the  GC system of `normal', old
ellipticals, as expected and as found by Paper I. The LF of the red GC system
as a whole exhibits a power-law behavior with slope $\alpha \sim 1.7$ (as was
found in Paper I using the earlier WFPC2 data). Although there is an
indication of flattening out beyond $M_V \sim -5.8$, this does not seem to
constitute strong evidence for the expected disruption of the low-mass end of
the second-generation GCs.    
{\it However}, the effect of cluster disruption processes should show up
first in the central regions of galaxies since the disruption timescale of
GCs scales with galactocentric distance (e.g., Vesperini \& Heggie 1997; Fall
\& Zhang 2001). With the huge number of detected GC candidates in the {\it
  ACS\/} data, we are now in a position to produce GC LFs in different radial
intervals with adequate statistical significance. And indeed, as panels (c)
and (d) of Fig.~1 show, the LF of the outer 50\% of the red GCs reveals a 
power-law down to its 50\% completeness limit, whereas the LF of the inner
50\% of the red GCs {\it does\/} shows a turnover at $M_V \sim 6$, i.e.\ 
$\sim$\,1 mag fainter than the turnover of `old' GC systems.  

\begin{figure}
\centerline{
\psfig{figure=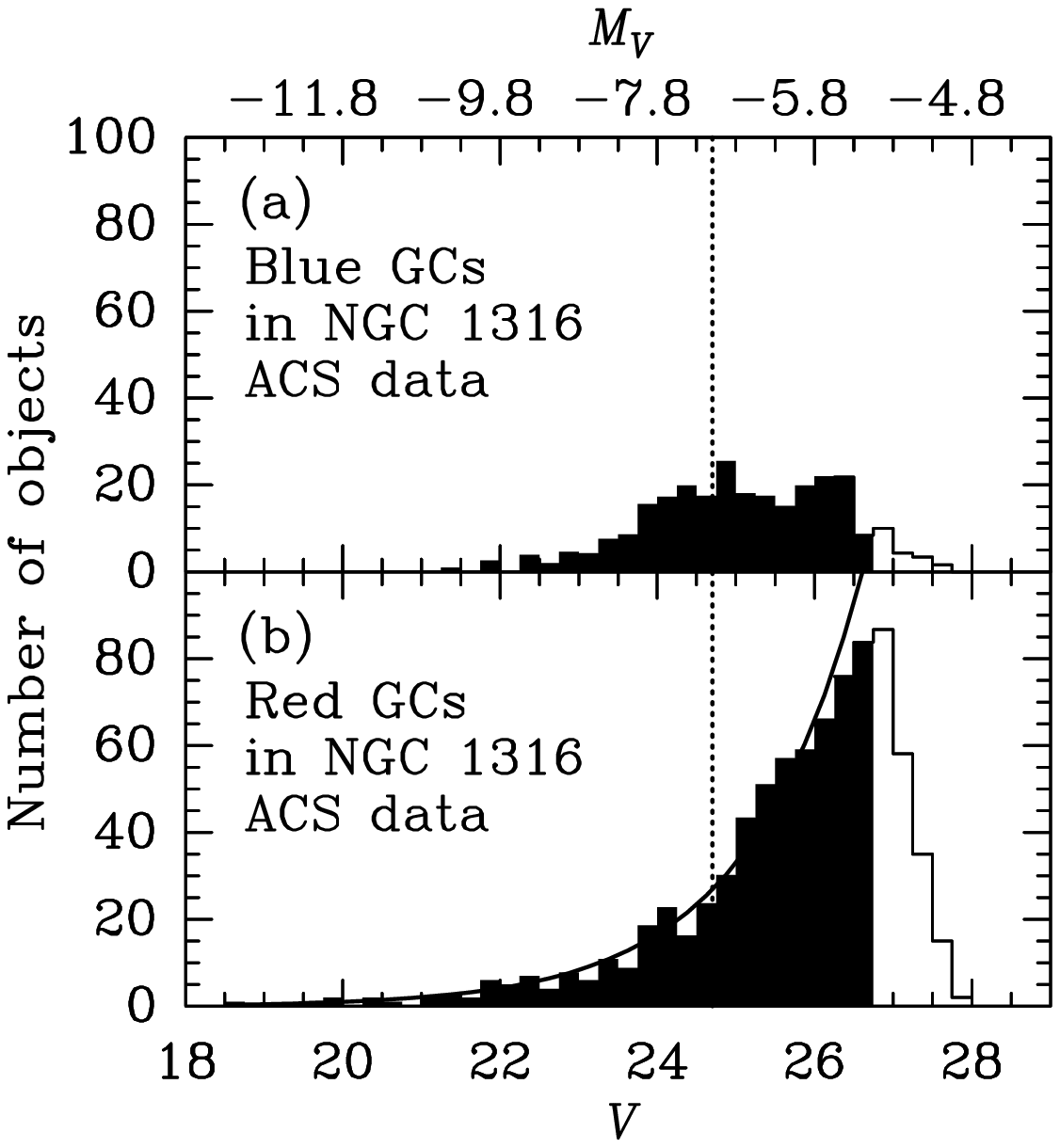,height=6.6cm}
\hspace*{1mm}
\psfig{figure=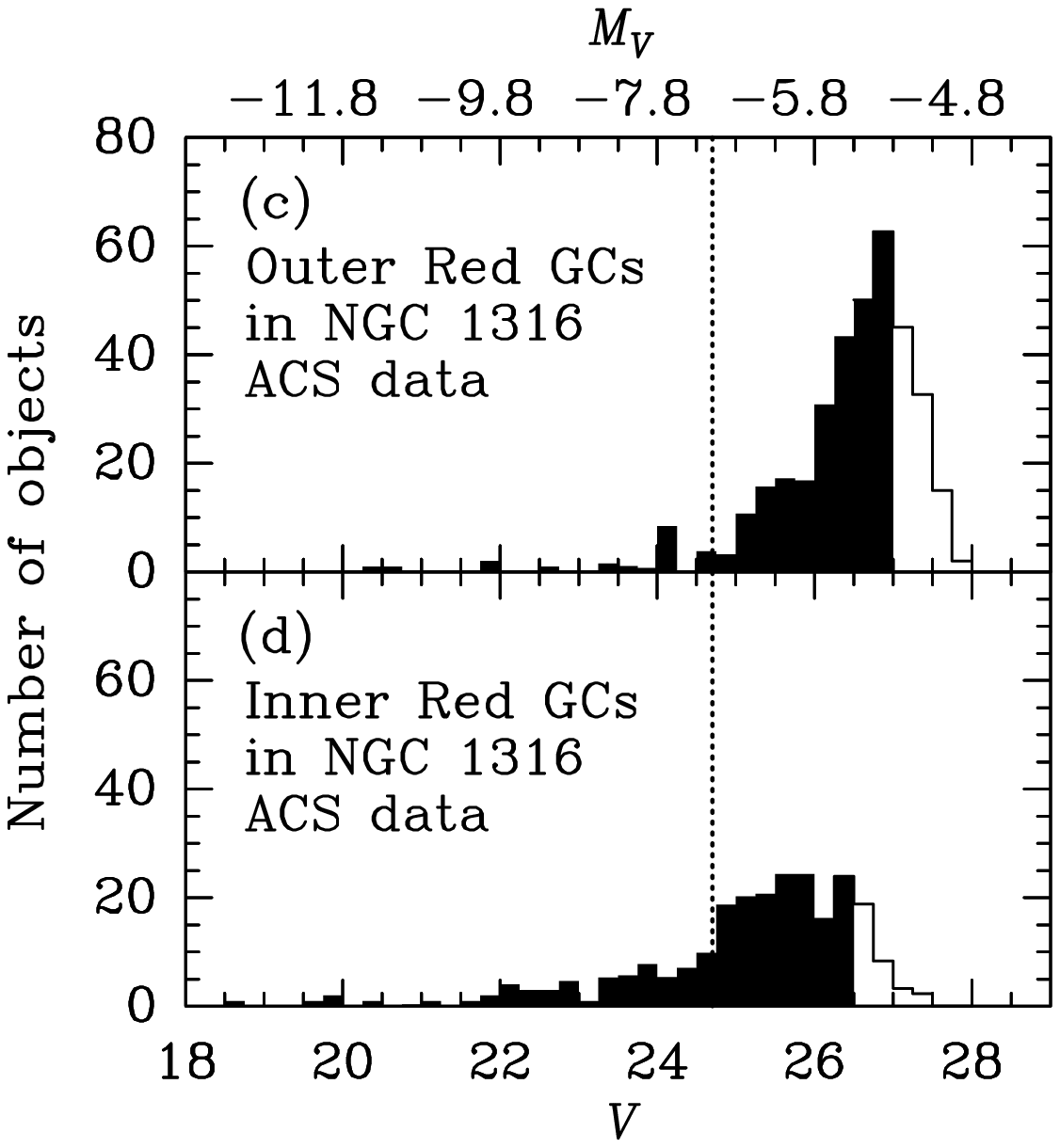,height=6.6cm}
}
\caption{$V$-band LFs of GC candidates from the
  {\it ACS\/} data. Panel (a): LF of the full `blue' subpopulation. Panel 
  (b): LF of the full `red' subpopulation. Panel (c): 
  LF of the outer 50\% of the `red' subpopulation. Panel (d): 
  LF of the inner 50\%\ of the `red' subpopulation. The histograms are
  filled for magnitude bins brighter than the {\it weighted mean\/} 50\%\
  completeness limit (this limit depends strongly on the background
  level), and open beyond it. The smooth curve in panel (b) is a
  power-law fit to the LF. The dotted vertical lines represent the
  predicted turnover magnitude for `old' GC systems. 
  }
\end{figure}

\paragraph{Acknowledgments.}
It is a pleasure to thank the workshop
  organizers for a {\it great\/} 
workshop that brought together researchers from a variety of fields
related to star cluster research. I thank my collaborators on this
project:\ Diane Karakla, Fran\c{c}ois Schweizer, and Brad Whitmore. Support
for HST Proposal number GO-9409 was provided by NASA through a grant from the
Space Telescope Science Institute, which is operated by the Association of
Universities for Research in Astronomy, Inc., under NASA contract
NAS5--26555.

\end{document}